%% file: paper_preprint.tex
\documentclass[11pt]{article}
\usepackage{epsfig}
\usepackage{amsmath}
\usepackage{hhline}
\usepackage{amssymb}
\usepackage{times}
\newlength{\dinwidth}
\newlength{\dinmargin}
\begin{document}  
% The rest
\newcommand{\pom}{{I\!\!P}}
\newcommand{\slowpi}{\pi_{\mathit{slow}}}
\newcommand{\fiidiii}{F_2^{D(3)}}
\newcommand{\fiidiiiarg}{\fiidiii\,(\beta,\,Q^2,\,x)}
\newcommand{\n}{1.19\pm 0.06 (stat.) \pm0.07 (syst.)}
\newcommand{\nz}{1.30\pm 0.08 (stat.)^{+0.08}_{-0.14} (syst.)}
\newcommand{\fiidiiiful}{F_2^{D(4)}\,(\beta,\,Q^2,\,x,\,t)}
\newcommand{\fiipom}{\tilde F_2^D}
\newcommand{\ALPHA}{1.10\pm0.03 (stat.) \pm0.04 (syst.)}
\newcommand{\ALPHAZ}{1.15\pm0.04 (stat.)^{+0.04}_{-0.07} (syst.)}
\newcommand{\fiipomarg}{\fiipom\,(\beta,\,Q^2)}
\newcommand{\pomflux}{f_{\pom / p}}
\newcommand{\nxpom}{1.19\pm 0.06 (stat.) \pm0.07 (syst.)}
\newcommand {\gapprox}
   {\raisebox{-0.7ex}{$\stackrel {\textstyle>}{\sim}$}}
\newcommand {\lapprox}
   {\raisebox{-0.7ex}{$\stackrel {\textstyle<}{\sim}$}}
\def\gsim{\,\lower.25ex\hbox{$\scriptstyle\sim$}\kern-1.30ex%
\raise 0.55ex\hbox{$\scriptstyle >$}\,}
\def\lsim{\,\lower.25ex\hbox{$\scriptstyle\sim$}\kern-1.30ex%
\raise 0.55ex\hbox{$\scriptstyle <$}\,}
\newcommand{\pomfluxarg}{f_{\pom / p}\,(x_\pom)}
\newcommand{\dsf}{\mbox{$F_2^{D(3)}$}}
\newcommand{\dsfva}{\mbox{$F_2^{D(3)}(\beta,Q^2,x_{I\!\!P})$}}
\newcommand{\dsfvb}{\mbox{$F_2^{D(3)}(\beta,Q^2,x)$}}
\newcommand{\dsfpom}{$F_2^{I\!\!P}$}
\newcommand{\gap}{\stackrel{>}{\sim}}
\newcommand{\lap}{\stackrel{<}{\sim}}
\newcommand{\fem}{$F_2^{em}$}
\newcommand{\tsnmp}{$\tilde{\sigma}_{NC}(e^{\mp})$}
\newcommand{\tsnm}{$\tilde{\sigma}_{NC}(e^-)$}
\newcommand{\tsnp}{$\tilde{\sigma}_{NC}(e^+)$}
\newcommand{\st}{$\star$}
\newcommand{\sst}{$\star \star$}
\newcommand{\ssst}{$\star \star \star$}
\newcommand{\sssst}{$\star \star \star \star$}
\newcommand{\tw}{\theta_W}
\newcommand{\sw}{\sin{\theta_W}}
\newcommand{\cw}{\cos{\theta_W}}
\newcommand{\sww}{\sin^2{\theta_W}}
\newcommand{\cww}{\cos^2{\theta_W}}
\newcommand{\trm}{m_{\perp}}
\newcommand{\trp}{p_{\perp}}
\newcommand{\trmm}{m_{\perp}^2}
\newcommand{\trpp}{p_{\perp}^2}
\newcommand{\alp}{\alpha_s}

\newcommand{\alps}{\alpha_s}
\newcommand{\sqrts}{$\sqrt{s}$}
\newcommand{\LO}{$O(\alpha_s^0)$}
\newcommand{\Oa}{$O(\alpha_s)$}
\newcommand{\Oaa}{$O(\alpha_s^2)$}
\newcommand{\PT}{p_{\perp}}
\newcommand{\JPSI}{J/\psi}
\newcommand{\sh}{\hat{s}}
\newcommand{\uh}{\hat{u}}
\newcommand{\MP}{m_{J/\psi}}
\newcommand{\PO}{I\!\!P}
\newcommand{\xbj}{x}
\newcommand{\xpom}{x_{\PO}}
\newcommand{\ttbs}{\char'134}
\newcommand{\xpomlo}{3\times10^{-4}}  
\newcommand{\xpomup}{0.05}  
\newcommand{\dgr}{^\circ}
\newcommand{\pbarnt}{\,\mbox{{\rm pb$^{-1}$}}}
\newcommand{\gev}{\,\mbox{GeV}}
\newcommand{\WBoson}{\mbox{$W$}}
\newcommand{\fbarn}{\,\mbox{{\rm fb}}}
\newcommand{\fbarnt}{\,\mbox{{\rm fb$^{-1}$}}}
%
% Some useful tex commands
%
\newcommand{\qsq}{\ensuremath{Q^2} }
\newcommand{\gevsq}{\ensuremath{\mathrm{GeV}^2} }
\newcommand{\et}{\ensuremath{E_t^*} }
\newcommand{\rap}{\ensuremath{\eta^*} }
\newcommand{\gp}{\ensuremath{\gamma^*}p }
\newcommand{\dsiget}{\ensuremath{{\rm d}\sigma_{ep}/{\rm d}E_t^*} }
\newcommand{\dsigrap}{\ensuremath{{\rm d}\sigma_{ep}/{\rm d}\eta^*} }

\begin{titlepage}
\begin{flushleft}
{\tt DESY 98-063 \hfill ISSN 0418-9833 \\ May 1998} \\
%{\bf Only for circulation within EW Group,referees,physics coord.}\\
%{\tt Draft 2.1, May 25, 1998 \\
%Editors : C. Diaconu, J. Meyer, C. Vall\'ee \\
%Referees : R. Eichler, A. Mehta\\
%Comments to editors and referees by Friday, 22/5/98 ,noon\\
%(J.Meyer has H1-nickname 'meyerj' or address joachim.meyer@desy.de !!!) 
\end{flushleft}
\vspace*{3.0cm}
\begin{center}\begin{LARGE}
{\bf  Observation of Events with an Isolated High Energy  Lepton and
Missing Transverse Momentum at HERA}

\vspace*{2.5cm}
H1 Collaboration \\
\vspace*{2.5cm}
\end{LARGE}
%================================abstract===================
%\vspace*{0.5cm}
{\bf Abstract}
\begin{quotation}
\noindent 
%%%%%%%%%%%%%%%%%%% ABSTRACT%%%%%%%%%%%%%%%%%%%%%
A search for
events with an imbalance in transverse momentum
and with isolated high \mbox{energy} leptons has been carried out
at the positron-proton collider HERA.
One event with an  
$e^-$ and five events with a  $\mu^{\pm}$ are found  together
 with evidence for 
undetected particles \mbox{carrying} transverse momentum.
Within the Standard Model the dominant origin of events with
  this kind of topology is the production 
of $W$ bosons with subsequent leptonic decay.
Three of the six events are within measurement errors 
found in a region of phase space likely to be populated by
 this process, while the remaining events show 
kinematic properties which are atypical of  all Standard Model processes considered.
\end{quotation}
\vspace*{5.0cm}
Submitted to European Physics Journal C
\end{center}

\cleardoublepage
\end{titlepage}
\input{h1auts.tex}
\\
\\
\input{h1inst.tex}
\newpage
\input{body.tex}

\section*{Acknowledgements}

We are grateful to the HERA machine group whose outstanding
efforts have made this experiment possible. 
We thank
the engineers and technicians for their work in constructing and now
maintaining the H1 detector, our funding agencies for 
financial support, the
DESY technical staff for continual assistance, 
and the DESY directorate for the
hospitality which they extend to the non-DESY 
members of the collaboration. We  thank E. Boos for 
help in  calculations of some reaction cross sections.

% the bibliography

\newpage
%\section*{Figures}
%Figures at the end of the paper if possible please
\input{figures1.tex}
\end{document}

%% file: h1auts.tex
%   H1AUTS  Author list by names, no. of authors  361
%           status: 02/04/98   17.17.24
 C.~Adloff$^{34}$,                %WUPP-ST                  Adloff              
 M.~Anderson$^{22}$,              %MANC-LEFT  10/97         Anderson            
 V.~Andreev$^{25}$,               %LPI -PD                  Andreev             
 B.~Andrieu$^{28}$,               %ECPL-PD                  Andrieu             
 V.~Arkadov$^{35}$,               %ZEUT-ST    10/96         Arkadov             
 C.~Arndt$^{11}$,                 %DESY-LEFT  10/97         Arndt               
 I.~Ayyaz$^{29}$,                 %PARI-ST       5/96       Ayyaz               
 A.~Babaev$^{24}$,                %ITEP-PD                  Babaev              
 J.~B\"ahr$^{35}$,                %ZEUT-PD                  Baehr               
 J.~B\'an$^{17}$
% ...

%% file: h1inst.tex
%     H1 Institutes as appearing on publications
 $ ^1$ I. Physikalisches Institut der RWTH, Aachen, Germany$^a$ \\
 $ ^2$ III. Physikalisches Institut der RWTH, Aachen, Germany$^a$ \\
 $ ^3$ School of Physics and Space Research, University of Birmingham,
       Birmingham, UK$^b$\\
 $ ^4$ Inter-University Institute for High Energies ULB-VUB, Brussels;
       Universitaire Instelling Antwerpen, Wilrijk; Belgium$^c$ \\
 $ ^5$ Rutherford Appleton Laboratory, Chilton, Didcot, UK$^b$ \\
 $ ^6$ Institute for Nuclear Physics, Cracow, Poland$^d$  \\
 $ ^7$ Physics Department and IIRPA,
       University of California, Davis, California, USA$^e$ \\
 $ ^8$ Institut f\"ur Physik, Universit\"at Dortmund, Dortmund,
       Germany$^a$\\
 $ ^{9}$ DSM/DAPNIA, CEA/Saclay, Gif-sur-Yvette, France \\
 $ ^{10}$ Department of Physics and Astronomy, University of Glasgow,
          Glasgow, UK$^b$ \\
 $ ^{11}$ DESY, Hamburg, Germany$^a$ \\
% $ ^{12}$ I. Institut f\"ur Experimentalphysik, Universit\"at Hamburg,
%          Hamburg, Germany$^a$  \\
 $ ^{12}$ II. Institut f\"ur Experimentalphysik, Universit\"at Hamburg,
          Hamburg, Germany$^a$  \\
 $ ^{13}$ Max-Planck-Institut f\"ur Kernphysik,
          Heidelberg, Germany$^a$ \\
 $ ^{14}$ Physikalisches Institut, Universit\"at Heidelberg,
          Heidelberg, Germany$^a$ \\
 $ ^{15}$ Institut f\"ur Hochenergiephysik, Universit\"at Heidelberg,
          Heidelberg, Germany$^a$ \\
 $ ^{16}$ Institut f\"ur experimentelle und angewandte Physik, 
          Universit\"at Kiel, Kiel, Germany$^a$ \\
 $ ^{17}$ Institute of Experimental Physics, Slovak Academy of
          Sciences, Ko\v{s}ice, Slovak Republic$^{f,j}$ \\
 $ ^{18}$ School of Physics and Chemistry, University of Lancaster,
          Lancaster, UK$^b$ \\
 $ ^{19}$ Department of Physics, University of Liverpool, Liverpool, UK$^b$ \\
 $ ^{20}$ Queen Mary and Westfield College, London, UK$^b$ \\
 $ ^{21}$ Physics Department, University of Lund, Lund, Sweden$^g$ \\
 $ ^{22}$ Department of Physics and Astronomy, 
          University of Manchester, Manchester, UK$^b$ \\
 $ ^{23}$ CPPM, Universit\'{e} d'Aix-Marseille~II,
          IN2P3-CNRS, Marseille, France \\
 $ ^{24}$ Institute for Theoretical and Experimental Physics,
          Moscow, Russia \\
 $ ^{25}$ Lebedev Physical Institute, Moscow, Russia$^{f,k}$ \\
 $ ^{26}$ Max-Planck-Institut f\"ur Physik, M\"unchen, Germany$^a$ \\
 $ ^{27}$ LAL, Universit\'{e} de Paris-Sud, IN2P3-CNRS, Orsay, France \\
 $ ^{28}$ LPNHE, Ecole Polytechnique, IN2P3-CNRS, Palaiseau, France \\
 $ ^{29}$ LPNHE, Universit\'{e}s Paris VI and VII, IN2P3-CNRS,
          Paris, France \\
 $ ^{30}$ Institute of  Physics, Academy of Sciences of the
          Czech Republic, Praha, Czech Republic$^{f,h}$ \\
 $ ^{31}$ Nuclear Center, Charles University, Praha, Czech Republic$^{f,h}$ \\
 $ ^{32}$ INFN Roma~1 and Dipartimento di Fisica,
          Universit\`a Roma~3, Roma, Italy \\
 $ ^{33}$ Paul Scherrer Institut, Villigen, Switzerland \\
 $ ^{34}$ Fachbereich Physik, Bergische Universit\"at Gesamthochschule
          Wuppertal, Wuppertal, Germany$^a$ \\
 $ ^{35}$ DESY, Institut f\"ur Hochenergiephysik, Zeuthen, Germany$^a$ \\
 $ ^{36}$ Institut f\"ur Teilchenphysik, ETH, Z\"urich, Switzerland$^i$ \\
 $ ^{37}$ Physik-Institut der Universit\"at Z\"urich,
          Z\"urich, Switzerland$^i$ \\
\smallskip
 $ ^{38}$ Institut f\"ur Physik, Humboldt-Universit\"at,
          Berlin, Germany$^a$ \\
 $ ^{39}$ Rechenzentrum, Bergische Universit\"at Gesamthochschule
          Wuppertal, Wuppertal, Germany$^a$ \\
 $ ^{40}$ Visitor from Yerevan Physics Institute, Armenia \\
 $^{41}$ 
Institut f\"ur Experimentelle Kernphysik, Universit\"at Karlsruhe,
Karlsruhe, Germany

%\smallskip
% $ ^{\dagger}$ Deceased \\
 
\bigskip
 $ ^a$ Supported by the Bundesministerium f\"ur Bildung, Wissenschaft,
        Forschung und Technologie, FRG,
        under contract numbers 7AC17P, 7AC47P, 7DO55P, 7HH17I, 7HH27P,
        7HD17P, 7HD27P, 7KI17I, 6MP17I and 7WT87P \\
 $ ^b$ Supported by the UK Particle Physics and Astronomy Research
       Council, and formerly by the UK Science and Engineering Research
       Council \\
 $ ^c$ Supported by FNRS-NFWO, IISN-IIKW \\
 $ ^d$ Partially supported by the Polish State Committee for Scientific 
       Research, grant no. 115/E-343/SPUB/P03/002/97 and
       grant no. 2P03B~055~13 \\
 $ ^e$ Supported in part by US~DOE grant DE~F603~91ER40674 \\
 $ ^f$ Supported by the Deutsche Forschungsgemeinschaft \\
 $ ^g$ Supported by the Swedish Natural Science Research Council \\
 $ ^h$ Supported by GA~\v{C}R  grant no. 202/96/0214,
       GA~AV~\v{C}R  grant no. A1010619 and GA~UK  grant no. 177 \\
 $ ^i$ Supported by the Swiss National Science Foundation \\
 $ ^j$ Supported by VEGA SR grant no. 2/1325/96 \\
 $ ^k$ Supported by Russian Foundation for Basic Researches 
       grant no. 96-02-00019 \\

%% file: body.tex
\section{Introduction}

 At the HERA collider the study of rare topologies in the final 
state of  electron\footnote{In this paper the term `electron' 
is used to describe generically electrons or positrons.}-proton  interactions
 provides a unique possibility 
to search for new phenomena. 
In this paper, a search for events with an imbalance in 
transverse momentum and isolated high energy leptons is reported. 
In 1994 the 
 first event with such a signature  ($ e^+ p \rightarrow \mu ^+ X $)
was observed \cite{theevent,h1rsusy} in the H1 detector.  
Within the Standard Model (SM), such topologies are expected mainly 
from the production of $W$ bosons with subsequent leptonic decay. 
A significant excess of events compared to SM
expectations might  reflect 
anomalous electroweak three-boson couplings \cite{wwg}, 
or production of new particles as expected for example  
in supersymmetric extensions of the SM \cite{kon}.

 The present analysis is based on positron (27.5\gev) - proton (820\gev) 
collisions recorded by the H1 experiment between 1994 and 1997. The total 
integrated luminosity is $36.5 \pm  1.1$~pb$^{-1}$, an order of magnitude 
higher than in the  earlier studies cited above.

\section{Data Analysis}

\subsection{Experimental Conditions}

A detailed description of the H1 detector can be found in \cite{H1detector}.
Here we only summarize those properties of components essential  
for this analysis\footnote{The origin of the coordinate system is the nominal 
$ep$ interaction point. The direction of the incoming proton defines 
the $z$-axis for the measurement of polar and azimuthal angles as well
as for transverse momenta. It also defines the forward region of the detector. 
The modulus of a vector $\vec{V}$ is denoted $V$.
}.
   
A central tracker consisting of drift chambers is used to measure the
charged particle trajectories and to determine the interaction vertex.
Particle transverse momenta are determined from the curvature of the 
trajectories in a magnetic field of 1.15 Tesla.

Electromagnetic and hadronic final state particles 
are absorbed in a highly segmented liquid argon calorimeter 
\cite{h1cal} covering the polar angular range $4 \dgr < \theta < 153 \dgr$.
The calorimeter is  5 to 8 interaction 
lengths deep depending on the polar angle of the particle.
Electromagnetic shower energies are measured with a precision of 
$\sigma (E)/E = 12 \% / \sqrt{E/\mathrm{GeV}} \oplus 1\%$ and
 hadronic shower energies with 
$\sigma (E)/E = 50 \% / \sqrt{E/\mathrm{GeV}} \oplus 2 \%$
\cite{cerntests} using software-based energy weighting techniques. 
The absolute energies are known to 3\% and 4\% respectively. 
The liquid argon calorimeter is complemented in the backward region
by a lead/scintillating-fibre calorimeter\footnote{This device was installed 
in 1995  replacing  a  lead-scintillator ``sandwich'' calorimeter.} 
\cite{spacal} covering
the range $155 \dgr < \theta < 178 \dgr$, and by positron and photon
taggers located downstream of the
interaction point in the positron beam direction. 

The calorimeter is surrounded by a superconducting coil 
and an iron yoke instrumented with streamer 
tubes. Leakage of hadronic showers outside the  calorimeter is measured 
by the analogue  charge sampling of the streamer tubes (Tail Catcher) 
with a resolution of $\sigma (E)/E = 100 \% / \sqrt{E/\mathrm{GeV}}$ . 
Tracks of 
penetrating charged particles, such as muons, escaping the calorimeter
 are reconstructed from 
their hit pattern in the streamer tubes with an efficiency greater than 80\%. 

The trigger condition for interactions involving high transverse energy is
derived from the liquid argon calorimeter signals.
 The trigger efficiency is greater than 95\% for events 
with a scattered electron which has energy
above 10 GeV, and greater than 85\% for events with 
a calorimetric missing transverse momentum  which is greater than 25 GeV.

\subsection{Event Selection}

 The selection procedure uses methods developed in the framework of
 H1 Charged Current (CC) analyses \cite{cc}. A search for an imbalance
in transverse momentum with isolated high energy leptons is based 
on two independent
% pre-requisites: 
requirements:
 
\begin{itemize}
 \item The calorimetric missing transverse momentum, $P^{calo}_{T}$, 
 must be greater 
  than 25 GeV. 

  $P^{calo}_{T}$ is defined as the transverse
  component of the vector sum of all energy deposits recorded 
  in the liquid argon calorimeter and the tail catcher. The cut value of
  25 GeV is chosen to minimize contributions from 
    Neutral Current (NC) deep-inelastic scattering (DIS), 
  photoproduction and photon-photon processes, as well as from non-$ep$ 
  background. 
 %It  also  ensures a high trigger
 % efficiency for events with penetrating charged particles like muons. 

 \item There must be  at least one well measured central 
  track with a polar angle greater than $10\dgr$ and a transverse momentum 
  greater than 10 GeV.

  The cut value of 10 GeV is designed to minimize the contribution from
  semi-leptonic decays of heavy mesons. Tracks fulfilling the above 
  requirements are referred to as {\it high-$P_{T}$ tracks} in the following.

\end{itemize}

%A further requirement is the existence of a reconstructed vertex in the
%interaction region. 
Non-$ep$ events due to  cosmic muons,
 halo muons or beam-gas interactions 
are removed by using a set of topological and timing 
requirements \cite{cc} and by demanding a reconstructed vertex in the
$ep$ interaction region.  
 Residual  NC events with  energy loss in the calorimeter cracks 
or in the beam-pipe are 
eliminated by excluding events with an identified scattered positron
under the following conditions: 
either the positron is balanced by 
the hadronic system in azimuth within $\Delta\phi = 5\dgr$,
or  the longitudinal momentum balance results in
 $\delta  > 45 \gev$. Here $\delta = \sum E_i(1-\cos\theta_i)$, 
where $E_i$  and $\theta_i$ denote the energy and polar angle 
of each detected final state particle.
For an event where only longitudinal momentum along
the proton direction (proton-remnant region) is undetected, one expects
$\delta = 2E_e = 55 \gev$, where $E_e$ is the energy of the incident
positron.
 
 The resulting event sample consists of 124 events. It will be referred to 
as the \emph{inclusive sample} in the following.
  
Within the inclusive sample the isolation of high-$P_{T}$ tracks
with respect  to jets or  other tracks 
in the event is quantified using:
\begin{itemize}

\item Their distance $D_{jet}$ to the closest hadron jet in the 
pseudorapidity-azimuth plane $\eta-\phi$, defined by
$D_{jet} = \sqrt{(\eta_{track}-\eta_{jet})^2 + (\phi_{track}-\phi_{jet})^2}$.  
For this purpose  jets are reconstructed using a cone algorithm \cite{cone}
with $R=1$ and $E_T^{min}=5$ GeV.
The hadronic nature of a jet is ensured by requiring  the fraction of the total
jet energy deposited in the electromagnetic calorimeter to be 
smaller than 0.9 and the jet radius
to be larger than 0.1. The jet radius is defined in the $\eta-\phi$ plane
as the energy-weighted average distance to the jet axis of all jet components.

\item Their distance $D_{track}$ to the closest  track in  $\eta-\phi$, defined
in the same  way as  $D_{jet}$. Here all tracks fitted to the 
interaction vertex with a polar angle 
greater than $5\dgr$ are considered.

\end{itemize}

In all of the 124 events at least one hadron jet is found.
Figure \ref{isolation} shows the correlation between $D_{track}$ 
and $D_{jet}$ for the high-$P_{T}$  tracks in the inclusive event sample.
In most cases the high-$P_{T}$ tracks are not isolated.
This is expected, since the bulk of
events are  CC interactions with the  high-$P_{T}$ track
located within or close to the  hadronic shower.
However, six high-$P_{T}$ tracks are found  in a region well 
separated from all other charged tracks and from hadron jets. 
They belong to six events each with one single isolated 
high-$P_{T}$ charged particle.    
 
\par
The nature of the six isolated high-$P_{T}$ particles is 
investigated by applying the following lepton identification criteria:
 
\begin{itemize}
  \item Electron candidate: 
        A calorimetric energy deposit larger than $ 5 \gev$, with more than 
    90\% of the energy located in the electromagnetic part 
    of the liquid argon calorimeter, has  
        to match geometrically the track.
  \item Muon candidate: The central track extrapolation must have a 
        total liquid argon energy smaller than 5 GeV in its vicinity, 
        and it has to match geometrically an instrumented iron signal 
        (muon system track or tail catcher energy deposit).
\end{itemize} 

All six isolated high-$P_{T}$ particles fulfill one of 
the lepton identification
criteria. One is an electron candidate and five are muon
candidates.

In summary, after application of an inclusive selection procedure
and lepton identification criteria, six events are found with 
calorimetric missing transverse momentum and an isolated high-$P_{T}$ 
lepton candidate, from which five are muon candidates and one is an electron 
candidate. No event is found with more than one high-$P_{T}$ lepton candidate.
The six events are described in detail in the next section.

\subsection{Event Properties}

% Displays of the six events (labelled ELECTRON and MUON-1 to -5) are shown 
%in figures \ref{thefourevents1} -- \ref{thefourevents6}.

Hereafter the six events with lepton candidates are labelled ELECTRON 
and MUON-1 to -5. Their event displays are shown in figure  \ref{thesixevents}.

\subsubsection* {Lepton Signatures}

The electron candidate shows an isolated track measured in the central 
tracker with a specific ionization, ${\rm d}E/{\rm d}x$, compatible with
the passage of a single particle.
The track is linked to an energy deposit ($E = 82 \gev,  E_T = 38 \gev$)
in the electromagnetic part of the calorimeter. The shower pattern 
is compatible with the  expectation 
 for a shower of electromagnetic origin.
The measured curvature of  the track is compatible with that expected for
 a negatively charged 
particle with a transverse momentum of $P_T = 38 \gev$.
The assignment of a positive charge is incompatible with the 
measurement at the 
level of 5 standard deviations.

In all five muon candidates  
the isolated track is measured in the central tracker, 
the calorimeter and the chambers of the instrumented iron system.
For all tracks the specific ionization,  as
measured in the central tracker, is consistent with
a single minimum ionizing particle traversing the chambers.  
The track is assigned a positive charge for MUON-1 and -2, 
and a negative charge for MUON-3 and -4. 
For MUON-5 the charge is undetermined. 
The central track is extrapolated through the
calorimeter and the coil to the instrumented iron system,
taking into account the effect of the magnetic field and the expected multiple
scattering and energy loss in the material along the path.
For all events the energy depositions in the calorimeter 
sampled over a path length of $\approx 7$ interaction lengths
are compatible in shape and magnitude
with those expected for a minimum ionizing particle.
In four of the five events the central track is linked successfully
to a track reconstructed in the muon chambers. 
In event MUON-5 the instrumented iron signals  line up with the 
extrapolated central track but are recorded near an edge of   
acceptance, preventing reconstruction of a track in the muon chambers. 
%The signals (hit multiplicity, energy deposition)
%recorded in the instrumented iron  are in all cases as expected
%for a single penetrating particle.

The isolated muon candidates penetrate in total more than 14 interaction 
lengths in the calorimeter and the instrumented iron.
In the momentum range of the muon candidates,
the probability that an isolated charged hadron would simulate a muon both in the 
calorimeter and the instrumented iron  is
% conservatively
estimated to be $ < 3 \cdot 10^{-3}$.
 
  In summary the signals seen in the central 
tracker, the calorimeter and the instrumented
 iron system support the
               interpretation of the electron candidate and the 5 muon 
               candidates as  $e^-$ and $\mu^{\pm}$, respectively.

\subsubsection* {Event Kinematics}

In all events 
a shower with hadronic signature is recorded in the calorimeter.
In event MUON-5 no charged particles are found to be correlated with the 
core of the high-$P_{T}$
hadronic jet. 
Event MUON-3  shows in addition 
to the high-$P_{T}$
muon  a low-$P_{T}$ positron measured in the central tracker 
and the calorimeter. No signals are seen in the backward calorimeter, 
the positron or the photon taggers, apart from an energy deposition of 3 GeV
registered for event MUON-4 at the edge of the acceptance in the innermost region of the backward 
calorimeter.

The event kinematics are quantified using the following variables:

\begin{itemize}
 \item $\vec{P}_{T}^{\ell}$ : Transverse momentum of the isolated lepton. 
 It is calculated using the calorimetric information for electrons and 
 the central tracker information for muons.
\item $\vec{P}^{X}_{T}$ :  Transverse momentum of the hadron system.
 $\vec{P}^{X}_{T}$ is defined as the transverse
 component of the vector sum of all energy deposits recorded 
in the liquid argon calorimeter and the tail catcher, except for those 
associated with  isolated leptons  (for MUON-3 both the isolated muon and
positron are excluded from the system $X$). 
\item $P_{\|}^{X}$ :
  Component of $\vec{P}^{X}_{T}$  parallel to the 
transverse direction of the isolated high-$P_{T}$ lepton.
 \item $P_{\bot}^{X}$ :
 Component of $\vec{P}^{X}_{T}$  perpendicular to the 
transverse direction of the isolated high-$P_{T}$ lepton
(the sign of $P_{\bot}^{X}$ is defined along a y-axis such that 
$\vec{P}_{T}^{\ell}$, y, z form
  a right handed coordinate system).
 \item $P_{z}^{X}$ :
 Longitudinal momentum of the hadron system.
 \item $E^{X}$ :
 Energy of the hadron system.
\item $\vec{P}_T^{miss}$ :  Missing total transverse momentum of the event.
 $\vec{P}_T^{miss} = - (\vec{P}_T^{lepton(s)}  + \vec{P}_T^{X})$ 
is defined from the vector sum of transverse momenta of 
all observed final states particles. 
 The quantity $ P_T^{miss}$ is identical to $P^{calo}_{T}$ unless the event
contains 
 isolated muons in the  final state. Note that in the latter case 
 the $P^{calo}_{T}$ criterion in the inclusive selection does not 
 automatically imply an imbalance in the total transverse momentum
 of the event.
\item $\delta$ : event balance in longitudinal 
 momentum as defined in section 2.2.
\item $M_T^{\ell\nu} = \sqrt{(P_T^{miss} +P_T^{\ell})^2 -
(\vec{P}_T^{miss}+ \vec{P}_T^{\ell})^2}$ : high-$P_T$ lepton-neutrino 
transverse mass,
where the measured missing transverse momentum $\vec{P}_T^{miss}$
is attributed to a hypothetical neutrino.  

\end{itemize}

 The determination of the global event variables $P_T^{miss}$, $\delta$
and $M_T^{\ell\nu}$
relies on the measurement of the lepton and hadron kinematic variables. 
In all events the dead material and energy-weighting correction factors
applied to the hadron system $X$ in the reconstruction were checked to
be compatible with the average corrections to NC hadron jets in the same
$P^{X}_{T}$ and polar angle domain.  
 For the muon events the measurement of the high-$P_{T}$ muon track by 
the central tracker was 
checked using a sample of NC events, selected such that the
positron traverses the same region of the chambers as the muons. 
In this sample the comparison of the positron track parameters with the 
calorimetric measurement   
shows no significant bias of the tracker response, and no smearing towards
unphysically
 high transverse momentum values. In the domain under consideration 
the  resolution
of the track momentum is measured to be 
$\sigma_{P_T}/P_T^2=7\times10^{-3}$ GeV$^{-1}$,
 which is in the same range as the 
errors quoted for the muon tracks.
% The probability for the scattered 
%positron to be assigned the wrong electric charge is found to be less than
%1\%.

 The kinematic parameters of the events are summarized 
in table \ref{kinematics}.

\input{table.events.tex}

For all events an imbalance in total transverse momentum, $P_T^{miss}$,
is observed. This imbalance is ascribed to unobserved particles 
carrying away transverse momentum. Apart from the events MUON-4 and MUON-5, 
a significant deficit is also seen in the longitudinal momentum balance 
quantified by $\delta$. 
 For the events MUON-4 and MUON-5, the muon momenta are high 
and only measured with moderate precision resulting in large errors 
for $P_T^{miss}$ and $\delta$.
Additional evidence for undetected particles carrying transverse momentum 
is provided by the acoplanarity observed in most events
(table \ref{kinematics})
between the high-$P_{T}$ lepton and the hadronic system $X$, quantified by
$\Delta\phi = \arctan(P_{\bot}^{X}/P_{\|}^{X})$.

 An imbalance in  transverse 
momentum may be faked as a result of  measurement uncertainties or 
  high energy proton remnants lost in the beam pipe.
In order to assess the significance of the observed imbalance in 
transverse momentum, we compare in  
figure \ref{ptmiss}  the correlation between $\Delta\phi$ and $P_T^{miss}$
for the six events  under discussion 
with that measured in  NC events, which are expected to
be intrinsically coplanar and balanced in $P_{T}$.
% NC events are selected in  a  region of phase space
%($ \theta^\ell,  P_T^\ell,  P_T^X$)
%  similar to that populated  by the six events.
In order to cover a similar kinematic region, NC events are selected  
with $10\dgr < \theta^e < 50\dgr$, and, 
either $P_T^e > 20$ GeV and $P_T^X > 5$ GeV for comparison with  the 
ELECTRON event, or $P_T^e > 10$ GeV and $P_T^X > 25$ GeV for comparison with 
the MUON events. In the latter case $P_T^{miss}$ 
is computed using the 
positron track parameters instead of using the calorimetric information.
When  compared to the  NC events,
 all candidates except MUON-1 have both high $\Delta\phi$ and high 
 $P_T^{miss}$.
The probability for a NC event to have  both $\Delta\phi$ and $P_T^{miss}$ 
values greater than those measured in a given candidate is
determined from a high statistics simulation to 
1\% for MUON-1 and less than 0.1\% for the other candidates.
 
\section{Discussion}

 In the following we discuss processes within the SM
which may yield events
 with missing transverse momentum and  with an
isolated high-$P_T$ lepton.
The rates  predicted for these processes
 are summarized in table \ref{tableback}.
%A comparison is performed
%of the observed event properties with those expected from 
%the dominant SM contributions.
 
 Predictions for all processes considered are obtained by applying the  
selection procedure of section 2.2, by requiring a
high-$P_{T}$ lepton with $D_{track}>0.5$ and $D_{jet}>1$, and
by taking into account trigger efficiencies.
%For kinematics determination (see 2.3)
Additional isolated leptons, if any,  are identified using the same criteria
            as for the high-$P_T$
 lepton but lowering the minimum  $P_T$  to 1 GeV.
 The SM rates quoted 
do not include 
events with more than one isolated lepton of the same generation 
            (lepton pair production) 
since  this topology does 
            not correspond to the topologies of the observed six events.
 
The Monte-Carlo simulations used for the predictions include a full
simulation of the H1 detector response.
 
\subsection{Standard Model Processes}

 \begin{itemize}
\item {$W$} Production:
At parton level a lepton-neutrino pair can be produced via the 
reactions 
$ e^+ q \rightarrow e^+ q' \ell \nu $ and 
$ e^+ q \rightarrow \overline{\nu}  q' \ell \nu $. The reaction 
$ e^+ q \rightarrow e^+ q' \ell \nu $ dominates by a factor 20. 
Its Feynman graphs are shown
in figure \ref{wgraphs}a - g.
The main  diagrams are those involving production of
a $W$ ($e^+ q \rightarrow e^+ W^{\pm} q'$) with subsequent leptonic decay 
$W^{\pm} \rightarrow \ell^{\pm} \nu$ (figure \ref{wgraphs}a-e). 
An additional source of $W$ production is due to resolved photon 
interactions. Here  the $W$ is produced by fusion of a quark from the proton 
             with  a quark from a long-lived hadronic fluctuation
of a photon emitted by the incident positron.
 
Positively and negatively charged $W$ bosons  
contribute with similar rates.
%  W \rightarrow \mathrm{e}^{\pm} \nu$  or 
%$ W \rightarrow \mu^{\pm} \nu)$
The expected total cross section 
%$\sigma (e p \rightarrow e W^{\pm} X) * BR
% (W^{\pm} \rightarrow \ell^{\pm} \nu )$
is about $ 60$ \fbarn~ per charge state and leptonic decay channel \cite{baur}.
 The cross section rises towards low
values of the transverse momentum $P_T^X$ of the recoil hadron system $X$,
and shows the Jacobian peak around the $W$ mass in the spectrum of the 
lepton-neutrino transverse mass $M_T^{\ell\nu}$. 
The resolved photon contribution dominates at very low $P_T^X$. 
The cascade decay $W\rightarrow \tau \nu ,
 \tau \rightarrow \mathrm{e},\mu$  contributes 
$\sim 15\%$ of the total accepted rates in the $e$ and $\mu$ channels.  
The scattered positron is
expected to be  observed in the 
calorimeters ($\theta~<~178~\dgr$) in 25\% of the events. 

%Additional
% contributions
% to the parton level
% diagrams (figure \ref{wgraphs})
% are
%the reaction $(e^+ q \rightarrow \overline{\nu} W  q, 
%W\rightarrow\ell \nu)$ ($\sim 5\%$ of the total accepted $W$ rate),
%and the cascade decay $W\rightarrow \tau \nu ,
% \tau \rightarrow \mathrm{e},\mu$   
%($\sim 15\%$ of the total accepted $W$ rate).  
%In the latter case the lepton--neutrino transverse mass 
%$M_T^{\ell\nu}$ may lie well below $M_W$.

$W$ production is studied quantitatively 
with the Monte-Carlo program EPVEC \cite{baur}, which
includes all leading-order
parton level diagrams for the reactions $ e^+ q \rightarrow e^+ q' \ell \nu $
and $ e^+ q \rightarrow \overline{\nu}  q' \ell \nu $
together with the resolved photon contribution. 
EPVEC is interfaced to the JETSET hadronization program \cite{JETSET}.
The cross sections for the leading-order diagrams were checked 
independently \cite{boos}.
 
For the electron channel the acceptance of the selection
procedure  is 33\%. For the muon channel it is only 10\%,
since here the cut in $P^{calo}_{T}$  acts
 as a $P_T^X$ cut.
%For the muon channel the acceptance grows steeply from a negligible value
%$at low $P_T^X$ ($P_T^X < 20 \gev$) up to a plateau at $\sim 55\%$ at high
%$P_T^X$ ($P_T^X > 40 \gev$). 
%This behaviour is due to the cut in $P^{calo}_{T}$ of the selection, 
%which in the muon channel acts as an effective $P_T^X$ cut. 
%Because of the dominance of the cross section at low $P_T^X$ the 
%overall acceptance is lower than in the electron channel and amounts to 10\%.

The rate estimates quoted in table \ref{tableback} include the following
 systematic effects:
\begin{itemize}
\item the uncertainty in the proton structure function 
      and the QCD scale at which it should be evaluated 
      (a variation from 100 GeV$^2$ to 10000 GeV$^2$ was assumed).
\item the uncertainty in the photon structure function entering the
      resolved photon contribution. Since this contribution favours low
      values of $P_T^X$
 only the prediction 
      in the electron channel is affected.
      
\item the uncertainty in the contribution from higher order QCD terms. 
      This was estimated from the effect
      of including parton showers \cite{showers} 
      prior to hadronization.  
\end{itemize}

It should be emphasized that it is presently not known whether an exact 
computation of higher order QCD terms may  change the 
prediction significantly.

\item  $Z$ production ($ e p \rightarrow e Z X,
 Z \rightarrow \ell^+\ell^-$): This process will only contribute
significantly for the $\tau$ - channel and only 
in case that one $\tau$ decays
leptonically and the other  hadronically.
The hadronic final states measured in the six events do not
show  characteristics of hadronic $\tau$ decays.
In particular the observed charged particle multiplicities 
exceed those corresponding to the dominant $\tau$ decays 
(1 or 3 charged particles).
The generator EPVEC \cite{baur} was used to estimate the expected
rate from this channel.

  \item  CC - DIS processes: These events are intrinsically unbalanced
   in transverse momentum  due
   to the final state 
   neutrino. An isolated lepton can only be produced due to 
   fluctuations in the hadronic final state.
  The generator DJANGO \cite{Django}
is used to simulate  this contribution.

  \item NC - DIS processes: These events have intrinsically an isolated 
    positron, but a significant $P^{calo}_{T}$ can only be produced 
    by fluctuations in the shower development 
   and the detector response. The rate of
    these events is  suppressed by the selection criteria 
    on $\Delta\phi$ and $\delta$ (see 2.2).
 The generator DJANGO \cite{Django}
is used to simulate  this contribution.

  \item Photoproduction processes:
   In this class, events with two jets of high transverse momenta
   could contribute if one jet is observed  in the detector
   and the other jet fluctuates into a single isolated hadron which is
   identified as a lepton. 
   Contributions are only expected in the muon channel since in the electron
   channel no significant $P^{calo}_{T}$ is produced. 
   A muon signature can be
   generated by a hadron when it either decays in flight into 
   a muon or traverses 
   the detector without hadronic interaction. 
   The expected rate resulting from such a process has been derived from the
   measured rate of events
   showing a jet of high transverse momentum ($P_T^{jet} > 25 \gev$)
   recoiling against a single isolated hadron track of $P_T > 10 \gev$. 
   Folding this  rate with the probability that a hadron is misidentified 
   as a muon (see section 2.3) provides the contribution to our sample.  
 
   A special class of  photoproduction processes is  the
   production of a heavy quark pair. In particular bottom quarks may 
   produce via semileptonic decays isolated leptons
   and (due to the escaping neutrino) a transverse momentum imbalance.
   The corresponding contribution is estimated using the generator 
   AROMA \cite{aroma}. Since preliminary 
measurements of H1 \cite{bcross} indicate that the 
   cross section for bottom production is too low in AROMA, we
   use for the rate estimate a fivefold increased cross section.

  \item Photon-photon interactions 
 ($\gamma  \gamma \rightarrow \ell^+ \ell^-$):
  Here we consider the interaction of a photon radiated
from the positron with 
  a photon radiated from the proton.
    For these processes
 one has to differentiate between $e^+e^-$,
   $\mu^+\mu^-,
   \tau^+\tau^-$ production in either elastic (no visible 
   hadron recoil) or inelastic processes.

   Since the  production of $e^+e^-$ does not lead to
    significant $P^{calo}_{T}$ this process does not contribute.
 
   The production of $\mu^+\mu^-$ contributes 
only when one muon of the 
    pair remains undetected. The process 
     contributes in two ways.
   Firstly, it can contribute to the  
       $e$ channel of our selection via
    elastic production. In this case the scattered 
   $e^+$ is registered in the detector and causes the  $P^{calo}_{T}$.
    The rate of these events is significantly suppressed by the selection
    cut on $\delta$ (see 2.2).
    Secondly,  this process can contribute to the  
       $\mu$ channel in our selection via
    inelastic production.
    Here the final state hadrons cause the  $P^{calo}_{T}$.
    In both cases the events are expected to have  only  small $P_T^{miss}$. 
    The muon which escapes detection in the beam pipe is 
    emitted preferentially in the backward direction and thus it
    cannot carry a significant transverse momentum.

      A small contribution to both electron and muon channels is due to 
    elastic $\tau^+ \tau^-$ production with one $\tau$
    decaying hadronically and the
    other one leptonically. Due to the small $\tau$ 
     mass  the decay particles are essentially collinear and the 
events are thus expected to have  small values of  
     $P_{\bot}^{X}$. In addition, as already mentioned for the $Z$ 
    contribution,
    the characteristics of the hadronic final state of the six  events are 
    inconsistent with this process.
 
   The photon-photon processes are simulated with the LPAIR 
 \cite{LPAIR} generator.
A test of this generator has been performed by analysing 
 $e^+e^-$ pair production.
In this test, events are selected requiring at least two visible isolated
electrons of opposite charge with $P_T$  greater than 10 GeV and 5 GeV 
respectively. The NC and QED Compton background 
processes are reduced to a negligible 
level by demanding in addition either that a third electron be visible in the 
detector, or that 
the highest $P_T$  electron  be of negative 
charge and that $\delta < 45$ GeV. In the high $P_T$ range under consideration 17 (quasi-)elastic
events ($P_T^X < $ 2 GeV) and 8 inelastic events ($P_T^X > $ 2 GeV)
 are observed, 
compared to expectations of 16.9 and 3.5 respectively. In both elastic and 
inelastic channels the measured kinematic distributions are compatible with
the predictions. With  the limited statistics of the test sample 
in mind, the
errors quoted in table 2 for the contribution of
the photon-photon process conservatively
allow the inelastic
contribution to vary within a factor 2.

\item 
Drell-Yan pairs from resolved photon-proton interactions 
($q\overline{q} \rightarrow \ell^+ \ell^-$):
Drell-Yan pairs have intrinsically low transverse momenta and thus cannot 
accommodate a hadronic system at high $P_T^X$. 
Therefore only $\tau^+\tau^-$ pairs could contribute, 
when one $\tau$ decays leptonically and the other  hadronically.
In order to fullfil the $P^{calo}_{T}$ cut the $\tau^+\tau^-$ pair 
must have an invariant mass higher than 50 \gev. In this mass range 
the cross section extrapolated from fixed target measurements
\cite{DY} 
is less than 0.1 fb, resulting in a negligible contribution. 
 
\item Halo muon from the proton beam:  The proton beam in HERA is 
      accompanied by a flux of halo muons generated by proton losses
      around the ring. 
%     At the highest possible muon energy of 820 GeV, 
%      the interaction of such a muon with a residual gas nucleon $N$ cannot 
%      generate particles with transverse momenta higher than 
%      $\sqrt{s_{\mu N}}/2 \approx 20$ GeV. 
%      In this process also $\delta$ cannot exceed $m_N \approx 1$ GeV, 
%      which is well below the values measured in the six events.
      The kinematics of the
  scattering of such a halo muon on a residual gas nucleon 
     restricts the  
  transverse momenta and $\delta$ to be below the values
     observed in the six events. 

\end{itemize}
 
 Table \ref{tableback} indicates that  
  $W$ production constitutes the largest contribution 
 followed by
the NC process (for $e^+$ events) and the photon-photon
process (for $\mu^{\pm}$ events). 
All other processes considered contribute negligibly.

 \input{table.rates.tex}

%\clearpage  
 \subsection { Comparison of event properties to SM expectation}
 The  one $e^-$  event found is  compatible in rate with the expectation from
 $W$ production, while other processes contribute negligibly.
 In the $\mu$ channel the  5 events found have to be compared with
an expectation of  $0.8 \pm 0.2$
 events, in which $W$ production and photon-photon processes dominate.

The measured kinematic parameters $P_T^X$  and $M_T^{\ell\nu}$
  of the six events are 
       compared in figure~\ref{mtpt}
 with the distributions in these variables 
predicted for $W$ production 
 and photon-photon processes. As already discussed  $W$ events accumulate
       at $M_T^{\ell\nu}$  values close to the $W$ mass and at low or
 intermediate $P_T^X$ values,
       whereas photon-photon events concentrate at both low $P_T^X$ and low
   $M_T^{\ell\nu}$
       due to the intrinsic transverse momentum balance of this process.

In the ELECTRON and the MUON-3 events the hadronic recoil systems 
have  relatively small transverse momenta $P_T^X$ 
and the $M_T^{\ell\nu}$ values 
show up in the Jacobian peak close to the $W$ mass. 
If in the MUON-3 event  the observed positron is identified with the scattered 
positron  the invariant mass of the muon-neutrino 
system can be calculated.
 The result is $M_{\mu\nu} = 82^{+19}_{-12}$ GeV in agreement with 
the $W$ mass. Both these events show kinematic properties  
typical for $W$ events.

Event MUON-5 also lies at moderate $P_T^X$ and shows a transverse mass
 $M_T^{\ell\nu}$
    which, though measured at a high value, is compatible within 2 standard 
    deviations with the $W$ interpretation.

%Although the value of $M_T^{\ell\nu}$ measured for event MUON-5 is
%beyond those expected for the W process, the 2 sigma limit reaches 
% into the region populated by 
%W events. Since also the value of  $P_T^X$ is not too high, this event 
%has to be considered as W candidate too.

Figure~\ref{mtpt} shows that the other three muon events are found
in regions of phase space where the predicted SM rate is low.
 Events MUON-2 and MUON-4 show large values of 
$P_T^X$, and event MUON-1 shows a  small value of $M_T^{\ell\nu}$,
  both of which are 
unlikely to occur in the $W$ or 
photon-photon interpretations.

%%%%%%%%%%%%%%%%%%%%%%%%%%%%%%%%%%%%%%%%%%%%%%%%%%%%%%%%%%%%%5

\section{Summary}
In positron-proton scattering at HERA,   events have  been selected
requiring an imbalance in calorimetric transverse momentum 
and the presence of a high-$P_T$ charged particle.
While the majority of these events have topologies as expected from 
deep-inelastic charged current interactions, 
six events show the prominent signature of 
an isolated high-$P_T$ lepton (0 $e^+$, 1 $e^-$,
 2 $\mu^+$, 2 $\mu^-$ and 1 $\mu$ of
undetermined charge) together with evidence for undetected 
particles carrying transverse momentum.
 
%Within the inclusive selection procedure
 The total yield of events
 with one isolated high-$P_T$ lepton which is 
 expected from SM 
processes is $2.4 \pm 0.5$ in the $e^{\pm}$ channel and $0.8 \pm 0.2$ 
in the $\mu^{\pm}$ channel. The main contribution is due to $W$ production,
estimated in leading order to be
$1.7 \pm 0.5$ and $0.5 \pm 0.1$ events respectively. 
The electron event and one of the muon events are found
  in a region of phase space likely to be
populated by $W$ production.
 Another muon event can, within its large measurement errors, also 
be accommodated in a $W$ interpretation.
  The kinematic properties of the remaining three muon events,
together with the overall rate excess in the muon
channel, disfavour an  interpretation of these events
within the SM processes considered.

%% file: table.events.tex
\begin{table}
\begin{center}                
 \begin {tabular} {|l|c|c|c|c|c|c|}
 \hline  
 & {\bf ELECTRON} & {\bf MUON-1} & {\bf MUON-2} & {\bf MUON-3} *) &
{\bf MUON-4} & {\bf MUON-5} \\ 
 \hline

%{\bf Lep }&&&&&&\\     &&&&&&\\      
\multicolumn{7}{|l|}{}\\
\multicolumn{7}{|c|}{\bf The isolated high-$P_T$ lepton }\\
\multicolumn{7}{|l|}{}\\
\hline &&&&&&\\

Charge &Neg.( $ 5\sigma$) &  Pos.( $ 4\sigma$)&Pos.( $ 4\sigma$)&Neg.( $ 4\sigma$)&Neg.( $ 2\sigma$)&unmeasured\\  &&& &&&\\           

  $P_T^l$&$37.6^{+1.3}_{-1.3}$  &  $23.4^{+7.5}_{-5.5} $ &$28.0^{+8.7}_{-5.4}$&
$38.6^{+12.0}_{-7.4}$&$81.5^{+75.2}_{-26.4}$&$ > 44$
\\ &&& &&&\\

 $\theta^l$ & $27.3\pm 0.2$&  $ 46.2 \pm 0.1 $ &$28.9\pm 0.1$& $35.5 \pm 0.1$&$28.5 \pm 0.1 $ &$31.0\pm 0.1$\\
&&& &&&\\
 \hline

%{\bf Had }&&&&&&\\&&&&&&\\
\multicolumn{7}{|l|}{}\\
\multicolumn{7}{|c|}{\bf The hadronic system }\\
\multicolumn{7}{|l|}{}\\
\hline &&&&&&\\

$P_T^X$&$8.0\pm0.8$&$42.2\pm3.8$&$67.4\pm5.4$&$27.4\pm2.7$&$59.3\pm5.9$&$30.0\pm3.0$\\ &&& &&&\\

$P_{\|}^X$ &$-7.2\pm0.8$&$-42.1\pm3.8$&$-61.9\pm4.9$&$-12.5\pm2.1$&$-57.0\pm5.5$&$-28.6\pm3.1$\\ &&& &&&\\

$P_{\bot}^X$ &$-3.4\pm0.9$&$-2.7\pm1.8$&$26.8\pm2.7$&$-24.3\pm2.5$&$-16.3\pm3.2$&$-9.1\pm2.3$\\ &&& &&&\\

$P_z^X$ &$79.9\pm4.4$&$153.1\pm9.1$&$247.0\pm18.9$&$183.7\pm13.6$&$118.9\pm12.1$&$145.4\pm8.2$\\ &&& &&&\\

$E^X$ &$81.1\pm4.5$&$162.0\pm10.0$&$256.9\pm19.5$&$186.8\pm14.0$&$141.7\pm13.7$&$154.8\pm9.1$\\ &&& &&&\\

%Mass (\gev) &$11.8\pm4.1$&$32.1\pm9.2$&$20.3\pm7.5$&$19.5\pm13.2$&$49.1\pm6.6$&$44.0\pm5.3$\\
%  &&&&&&\\

\hline

%{\bf GP } &&&&&&\\ &&&&&&\\
\multicolumn{7}{|l|}{}\\
\multicolumn{7}{|c|}{\bf Global properties }\\
\multicolumn{7}{|l|}{}\\
\hline &&&&&&\\

 $ P_T^{miss}$ &$30.6\pm 1.5$& $18.9^{+6.6}_{-8.3}$ &$43.2^{+6.1}_{-7.7}$&$42.1^{+10.1}_{-5.9}$&$29.4^{+71.8}_{-13.9}$&$>18$\\ &&& &&&\\

  $\delta $&$10.4\pm 0.7$
 & $18.9^{+3.9}_{-3.2} $&$17.1^{+2.5}_{-1.7}$ &$26.9^{+4.2}_{-2.9}$&$43.5^{+19.3}_{-7.2}$&$>22$\\ &&& &&&\\

 $M_T^{l\nu}$ &$67.7\pm 2.7$&$3.0^{+1.5}_{-0.9}$ & $22.8^{+6.7}_{-4.2}$&$75.8^{+23.0}_{-14.0}$
&$94^{+157}_{-54}$&$>54$\\
 &&&&&&\\

% $P$ & 0.001 & 0.05  & 0.001 & 0.001 & 0.001 & 00.001 \\
% &&&&&&\\
\hline
\multicolumn{7}{|l|}{}\\
\multicolumn{7}{|c|}{*) Positron in MUON-3 :}\\
%\multicolumn{7}{|l|}{}\\
\multicolumn{7}{|c|} { $P_T^e = 6.7\pm 0.4$ ,
 $P_{\|}^e = 6.1 \pm 0.4$,
 $P_{\bot}^e = -2.8\pm 0.2$ , $P_z^e = -3.7 \pm 0.2 $}  \\ 
\multicolumn{7}{|c|}{}\\
\hline
\end {tabular}
 \end{center}
\caption[]{Reconstructed event kinematics (see text). Energies, momenta and
masses are given in GeV and angles in degrees. For the  charge of the
high-$P_T$ lepton the significance of the determination is  given.
In case of event MUON-5 $2 \sigma$ limits are quoted for the muon momentum 
and derived quantities.}
% The value of $\delta$ quoted for MUON-4 does not
%take into account the energy seen in the backward calorimeter (see text).}
\label{kinematics}
\end{table}

%% file: table.rates.tex
\begin{table}[h]
\begin{center}
\begin{tabular}{|c|c|c|} \hline
 &  Electron Channel &  Muon Channel\\
 \hline \hline
Data & 0 $e^+$, 1 $e^-$ & 5\\ \hline \hline
$W$ production & $ 1.65 \pm 0.47$  &$  0.53 \pm 0.11$ \\ \hline
$Z$ production &$0.01 \pm 0.01 $ & $ 0.01 \pm 0.01$\\ \hline
CC - DIS & $0.02 \pm 0.01$ &  $ 0.01 \pm 0.01$ \\ \hline
NC - DIS & $ 0.51 \pm 0.10$ $e^+$, $0.02 \pm 0.01$ $e^-$  & $ 0.09 \pm 0.06 $ 
\\ \hline
Photoproduction &   $<0.02 $    & $<0.02$ \\ \hline
Heavy Quarks &  $<0.04$  & $<0.04$\\ \hline
Photon-Photon & $ 0.09\pm 0.03$ $e^+$, $0.04 \pm 0.01$ $e^-$    & $ 0.14
 ^{+0.14}_{-0.07} $
 \\
 \hline
\end{tabular}
\\
\caption{Observed and  predicted event rates.
The  limits given correspond to 95\% confidence level.
Unless stated otherwise the quoted numbers refer to the summed
production of  both lepton charged states. }
\label{tableback}
\end{center}
\end{table}

%% file: figures1.tex
%%%%%%%%%%%%%%%%%%%%%%%%%%%%%%%%%%%%%%%%%%%%%%%%%%%%%%%%%
\begin{figure}[htb]
\centering 
\epsfig{file=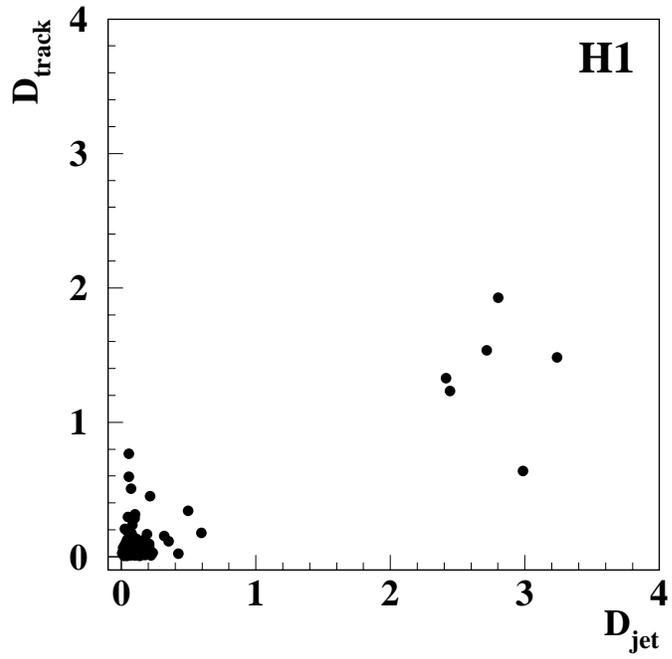,width=10.cm}
\caption{ Correlation between the distances $D_{jet}$ and $D_{track}$ 
(see text)
to the closest hadronic jet and track, for 
all high-$P_T$ tracks in the inclusive event sample.}
\label{isolation} 
\end{figure}
%%%%%%%%%%%%%%%%%%%%%%%%%%%%%%%%%%%%%%%%%%%%%%%%%%%%%%%%%%
\begin{figure}[htb]
\centering 
\epsfig{file=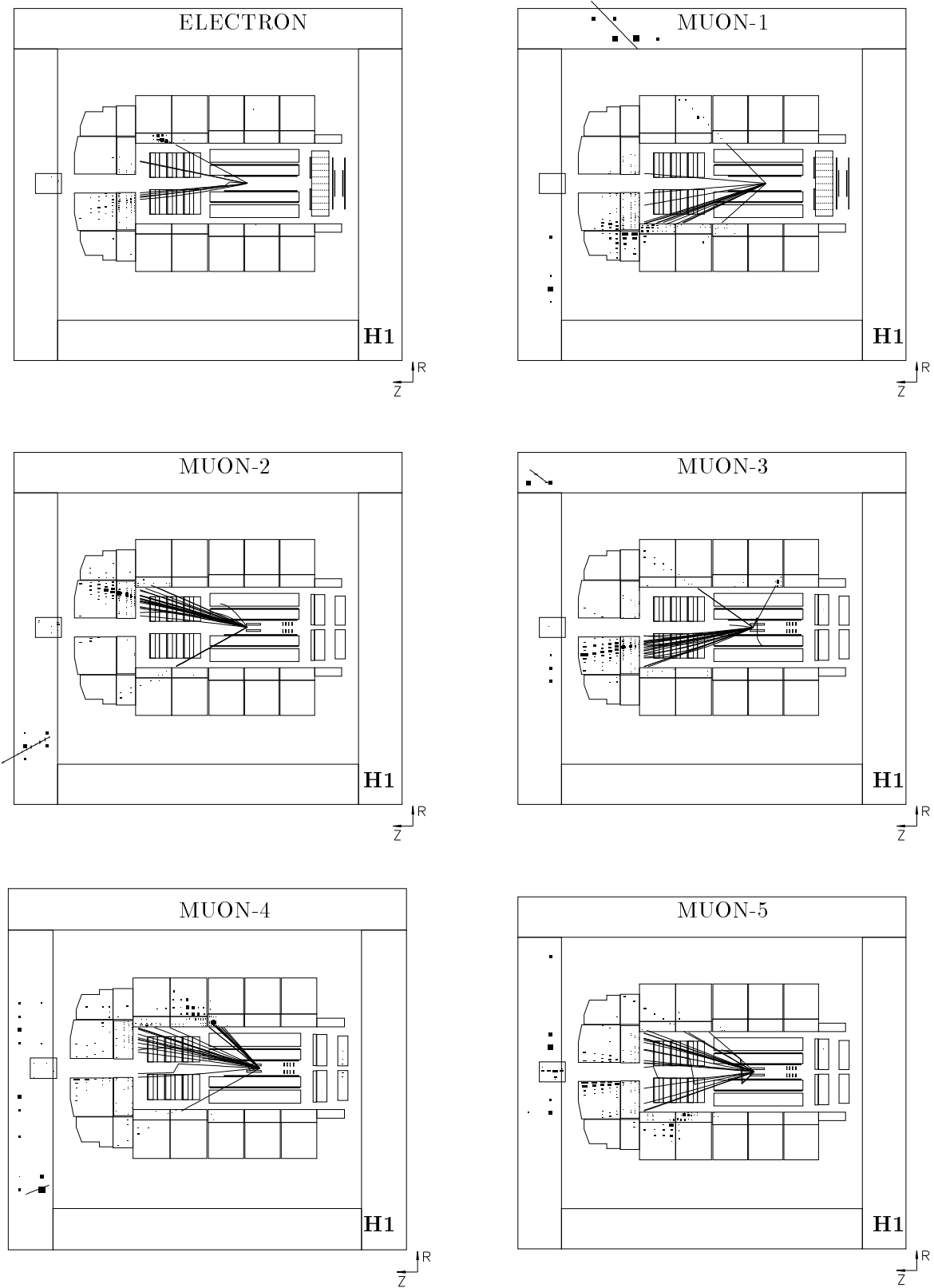,
bbllx=30pt,bblly=170pt,bburx=520pt,bbury=804pt,width=17cm} 
\vspace{0cm} 
\caption{Displays of the six events in the $R-z$ view. Indicated are
  the reconstructed tracks and the energy depositions in the calorimeters.
  The HERA positrons and protons enter the detector from the left and right,
  respectively.}
   \label{thesixevents}

\end{figure}

%%%%%%%%%%%%%%%%%%%%%%%%%%%%%%%%%%%%%%%%%%%%%%%%%%%%%%%%%%%%%%%%%%%%%
\begin{figure}[htb]
\centering 
\epsfig{file=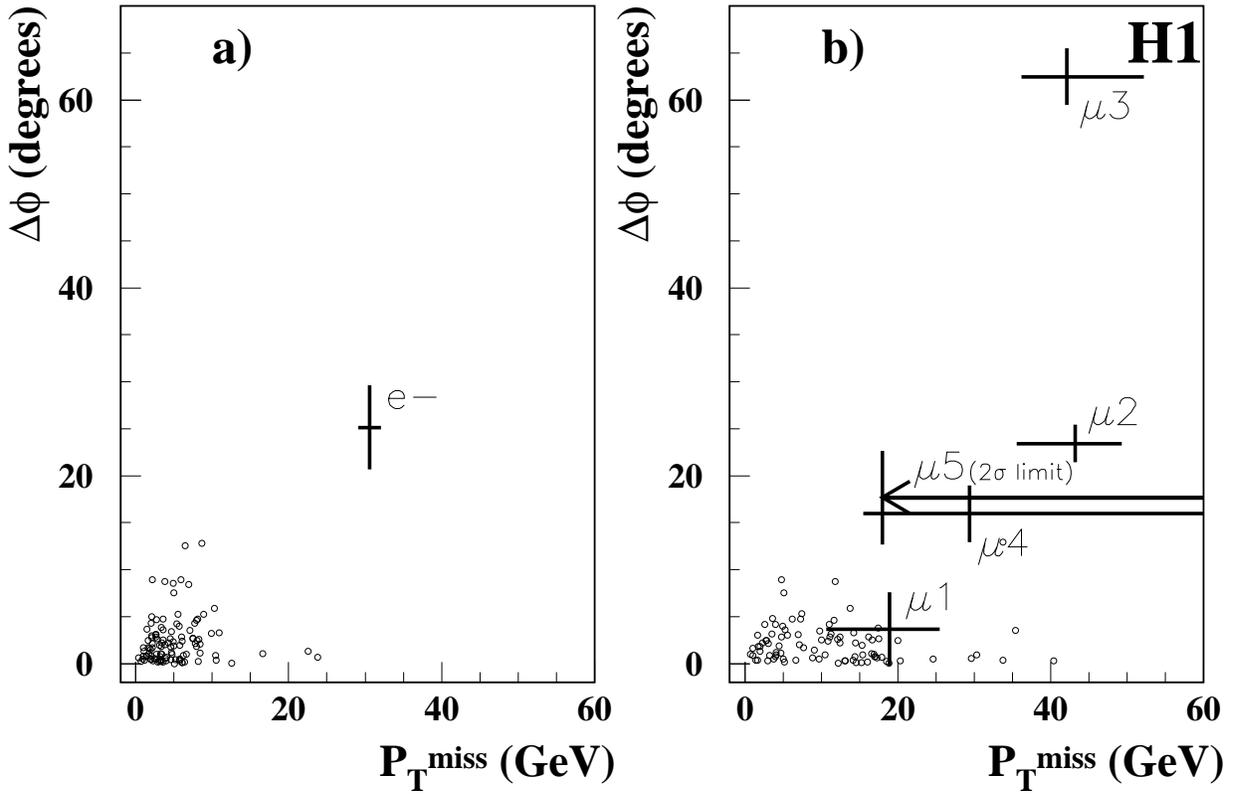, width=18.cm}
\caption{ Distribution of events in $P_T^{miss}$ and 
azimuthal acoplanarity $\Delta\phi$: a) electron channel; b) muon channel.
The six events are displayed with their individual measurement errors.
For comparison the open circles show the
 distributions of Neutral Current events 
(see text), 
with $P_T^{miss}$ computed in b) from the positron track 
instead of from the calorimeter shower.}
\label{ptmiss} 
\end{figure}
%
%%%%%%%%%%%%%%%%%%%%%%%%%%%%%%%%%%%%%%%%%%%%%%%%%%%%%%%%%%%%%%%%%%%%%%%%
\begin{figure}[htb]
\centering 
\epsfig{file=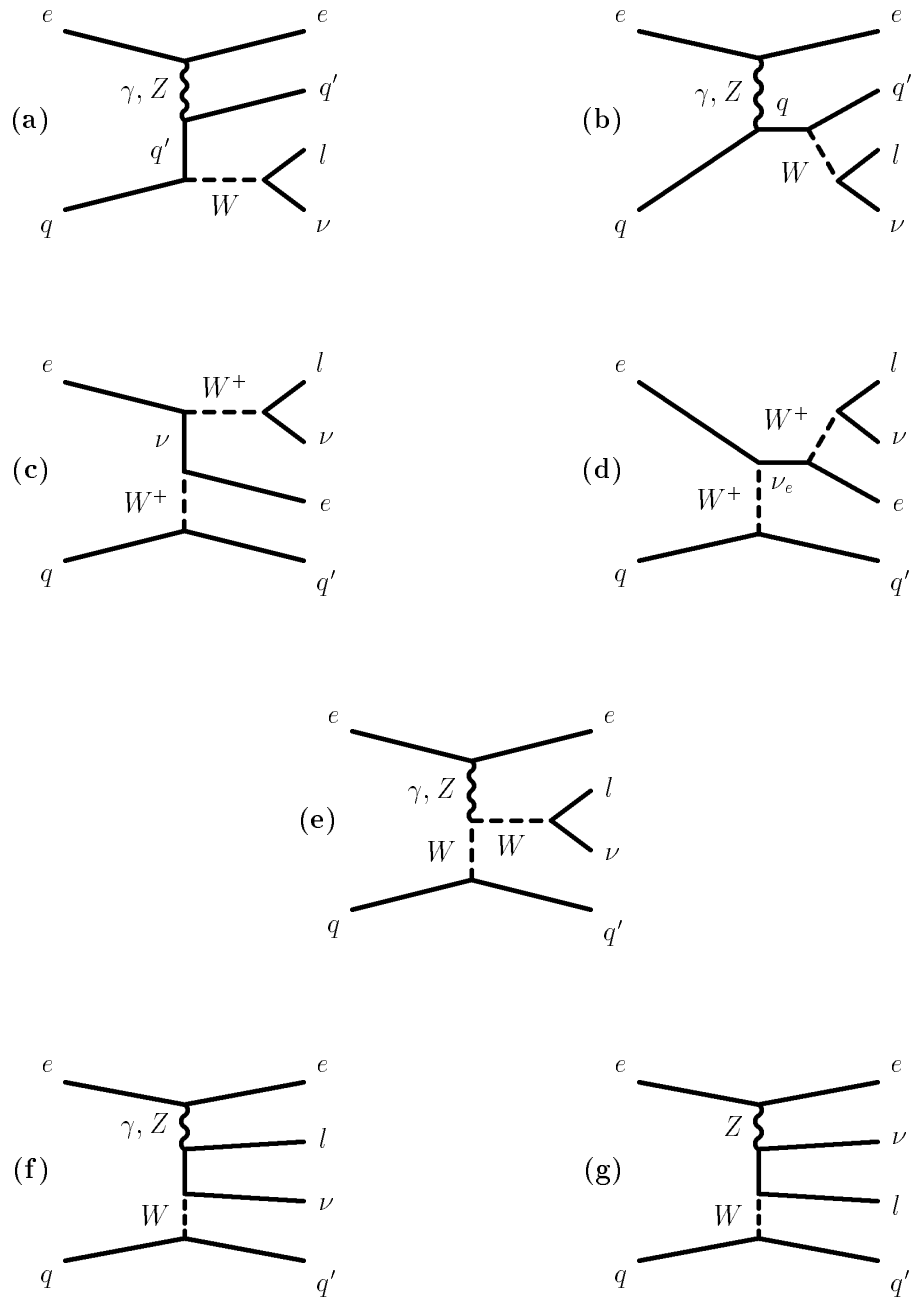,
bbllx=60pt,bblly=157pt,bburx=500pt,bbury=724pt,width=14cm} 
\caption{ Feynman diagrams for the process 
 $ e q \rightarrow e q' \ell \nu$}
\label{wgraphs} 
\end{figure}
%
%%%%%%%%%%%%%%%%%%%%%%%%%%%%%%%%%%%%%%%%%%%%%%%%%%%%%%%%%%%%%%%%%%%%%%%%%
%
%\begin{figure}[htb]
%   \centering 
% \epsfig{file=ettot_sum.1.5.eps, width=16.cm}
%   \caption{ Total scalar transverse momentum distribution of events with 
%isolated leptons observed (dots) in the extended kinematic domain 
%$P^{calo}_T > 15 \gev$ and $P_T^l > 2 \gev$, compared to expectations
%(histogram)
%from all standard model processes: 
%a) electron channel; b) muon channel.
%For events MUON-1 to -4 the horizontal bars indicate the measurement
%errors. 
%The dashed areas show the expected contributions from $W$ production 
%and the shaded areas the uncertainties on the overall predictions. }
%   \label{ettot} 
%\end{figure}
%%%%%%%%%%%%%%%%%%%%%%%%%%%%%%%%%%%%%%%%%%%%%%%%%%%%
\begin{figure}[htb]
   \centering 
 \epsfig{file=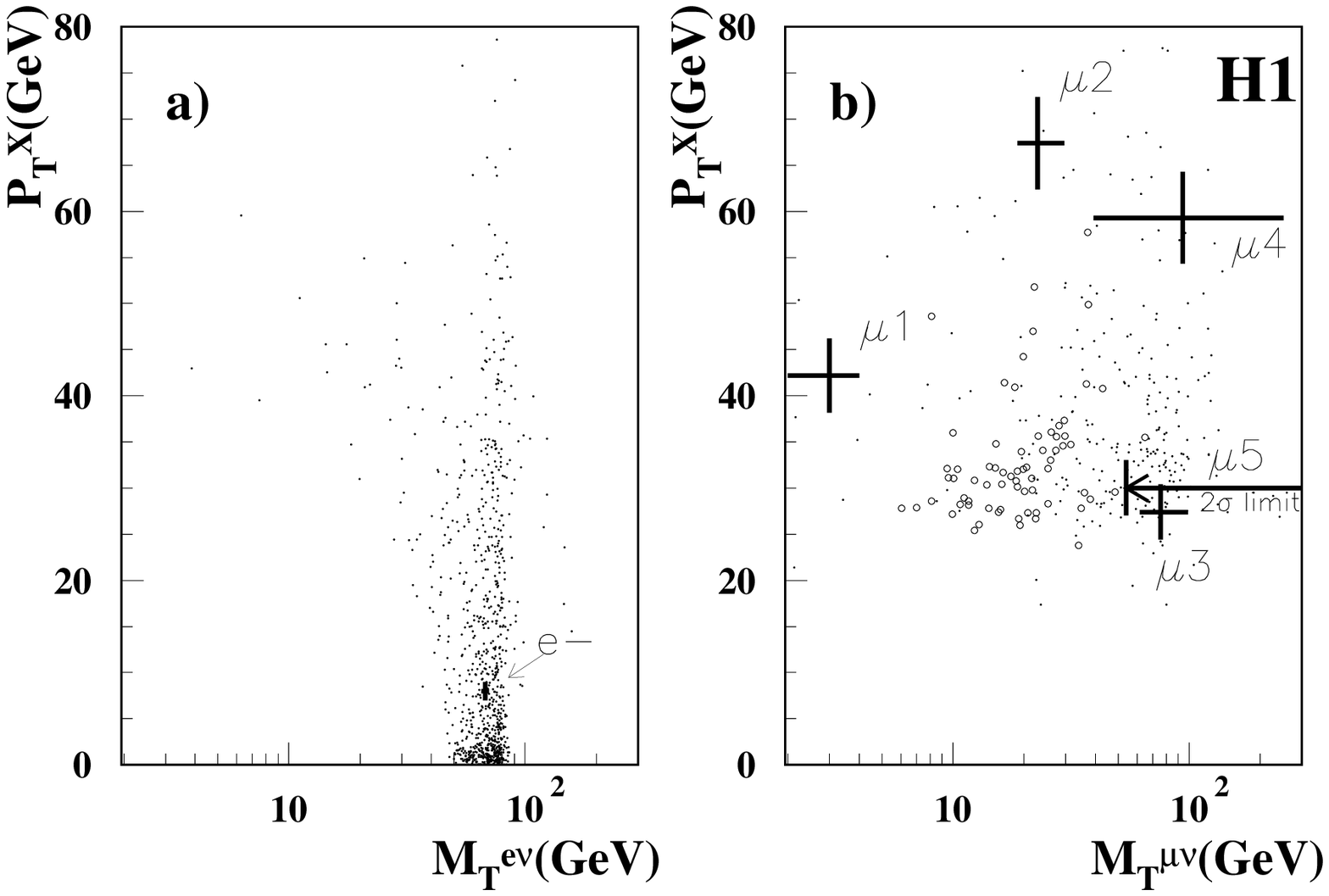, width=18.cm}
   \caption{ Distribution of  events 
 in  $P_T^{X}$ and $M_T^{l\nu}$: a) electron channel; b) muon channel.
% The ellipse contours correspond to the 1-sigma uncertainty on the
% measured parameters of the six observed events.  
 The crosses indicate the 1-sigma uncertainty on the
 measured kinematic parameters of the six observed events
 (for event MUON-5 the $2\sigma$ lower limit for $M_T^{l\nu}$ is shown).  
 The dominant SM  contributions (dots for $W$ 
 production, open circles 
 for photon-photon processes in the muon channel) are shown
 for an accumulated luminosity which is a factor 500 higher than in the data.
 In the muon channel no significant contribution is expected
 at $P_T^{X} < 25$ GeV because of the $P_T^{calo}$ cut 
 in the selection procedure.}
   \label{mtpt} 
\end{figure}
 
%%%%%%%%%%%%%%%%%%%%%%%%%%%%%%%%%%%%%%%%%%%%%%%%%%%%
%\begin{figure}[htb]
%   \centering 
% \epsfig{file=ettot.eps, width=16.cm}
%   \caption{ Total transverse energy distribution of events with 
%isolated leptons observed in the extended kinematical domain 
%$P^{calo}_T > 15 \gev$ and $P_T^l > 2 \gev$, compared to expectations
%from all standard model processes: 
%a) electron channel; b) muon channel.
%For events MUON-1 to -4 the horizontal bars indicate the measurement
%errors on $E_T^{tot}$. 
%The dashed areas show the expected contributions from $W$ production 
%and the shaded areas the uncertainties on the overall predictions. }
%   \label{ettot} 
%\end{figure}
 
%%%%%%%%%%%%%%%%%%%%%%%%%%%%%%%%%%%%%%%%%%%%%%%%%%%%%%%%%%%%%5

%% file: paper_preprint.bbl
\begin{thebibliography}{99}
\bibitem{theevent}
H1 Collaboration, T. Ahmed et al., DESY preprint 94-248 (1994).

\bibitem{h1rsusy} H1 Collaboration, S. Aid et al., Z.Phys.C71 (1996) 211.

\bibitem{wwg} M.N. Dubinin and H.S. Song, Phys.Rev. D57 (1998) 2927.

\bibitem{kon} T. Kobayashi, S. Kitamura and T. Kon, 
              Phys.Lett. B376 (1996) 227.

\bibitem{H1detector} H1 Collaboration, I. Abt et al., Nucl. Instr. and Meth.
A386 (1997) 310 and 348.

\bibitem{h1cal} H1 Calorimeter Group, B. Andrieu et al., 
Nucl. Instr. and Meth. A336 (1993) 460.

\bibitem{cerntests} H1 Calorimeter Group, B. Andrieu et al., 
Nucl. Instr. and Meth. A336 (1993) 499 and A350 (1994) 57.

\bibitem{spacal} H1 Spacal Group, R.D. Appuhn et al.,
Nucl. Instr. and Meth.  A386 (1997) 397.

\bibitem{cc}
H1 Collaboration, Z.Phys. C67 (1995) 565 and Phys.Lett. B379 (1996) 319. 

\bibitem{cone} 
M.H. Seymour, Z.Phys. C62 (1994) 127.

%\bibitem{DJANGO} DJANGO 6.2; G.A. Schuler and H. Spiesberger, Proc. of the 
%Workshop Physics at HERA, W. Buchm\"uller and G. Ingelman (Editors),
%(October 1991, DESY-Hamburg) Vol. 3 p. 1419.s
%\bibitem{gheisha}
% H. Fesefeldt, Th. Hamacher, J. Schug, Nucl. Instr. and Meth. A292  (1990) 279.
% 
\bibitem{baur}
U. Baur, J.A.M. Vermaseren, D. Zeppenfeld, Nucl. Phys. B375 (1992) 3.

\bibitem{JETSET}
T. Sj\"ostrand,  Comp. Phys. Com. 39 (1986) 347.

\bibitem{boos}
E. Boos, private communication.

\bibitem{showers}
T. Sj\"ostrand,  Comp. Phys. Com. 82 (1994) 74.

%\bibitem{gordon}
% L.E. Gordon and J.K. Storrow, Z. Phys. C63 (1994) 581.


\bibitem{Django}  DJANGO 6.2; G.A. Schuler and H. Spiesberger, Proc. of the 
Workshop Physics at HERA, W. Buchm\"uller and G. Ingelman (Editors),
(October 1991, DESY-Hamburg) Vol. 3 p. 1419.

\bibitem{aroma} G. Ingelman, J. Rathsman and G.A. Schuler, 
Comp. Phys. Comm 101 (1997) 135.

  \bibitem{bcross} U. Langenegger, Contribution to 
$33^{rd}$ Rencontres de Moriond : QCD and High Energy Hadronic Interactions,
Les Arcs (1998).

% \bibitem{h1hq} The prelm. H1 B-B-bar cross section has been used 
%                (5 * higher than AROMA !). Reference ??
 
\bibitem{LPAIR} S. Baranov et al., Proc. of Workshop Physics at HERA,
 W. Buchm\"uller and G. Ingelman (Editors), (October 1991, DESY-Hamburg) 
 Vol. 3, p. 1478;  J.A.M. Vermaseren, Nucl. Phys. B229 (1983) 347.

\bibitem{DY} J. Badier et al., Phys. Lett. 89B (1979) 145.
%\bibitem{comhep} E. Boos et al., preprint INP MSU 94-36/358 (1994), 
% hep-ph/9503280.
 


\end{thebibliography}
